\documentclass[useAMS,usenatbib]{mn2e}
\usepackage{graphicx}
\usepackage{lastpage}
\usepackage{txfonts}

\def\hii {H\,{\sc ii}}

\def\kms{km\,s$^{-1}$}

\title[Periodic methanol masers]{Discovery of four periodic methanol masers and updated light curve for a further one}

\author[M. Szymczak, P. Wolak and A. Bartkiewicz]
{M. Szymczak \thanks{E-mail: msz@astro.umk.pl},
P. Wolak
and A. Bartkiewicz
\\
Centre for Astronomy, Faculty of Physics, Astronomy and Informatics, Nicolaus Copernicus University,\\
 Grudziadzka 5, 87-100 Torun, Poland \\
}

\begin{document}


\pagerange{\pageref{firstpage}--\pageref{LastPage}} \pubyear{2014}

\maketitle

\label{firstpage}
\begin{abstract}
We report the discovery of 6.7\,GHz methanol maser periodic flares in four massive star forming regions
and the updated light curve for the known periodic source G22.357+0.066. The observations were carried out
with the Torun 32\,m radio telescope between June 2009 and April 2014. Flux density variations with
period of 120 to 245\,d were detected for some or all spectral features. A variability pattern with
a fast rise and relatively slow fall on time-scale of 30$-$60\,d dominated. A reverse pattern
was observed for some features of G22.357+0.066, while sinusoidal-like variations were detected 
in G25.411+0.105. A weak burst lasting $\sim$520\,d with the velocity drift of 0.24\,\kms\,yr$^{-1}$
occurred in G22.357+0.066. For three sources for which high resolution maps are available, 
we found that the features with periodic behaviour are separated by more than 500\,au from those without 
any periodicity. This suggests that the maser flares are not triggered by large-scale homogeneous variations 
in either the background seed photon flux or the luminosity of the exciting source and a mechanism which
is able to produce local changes in the pumping conditions is required.
\end{abstract}


\section{Introduction}
The 6.7\,GHz methanol maser transition has been detected from very many massive star-forming regions 
(\citealt{pandian07}; \citealt{caswell10, caswell11}; \citealt{green10, green12}; \citealt{szymczak12}). 
Since the early report of \citet{caswell95}, the temporal properties of methanol masers have been intensively studied. 
In particular, the phenomenon of periodic or regular variability at 6.7\,GHz in high-mass young stellar objects (HMYSOs) 
is a matter of active debate.

To date, periodic variations in the maser emission have been detected from a dozen HMYSOs (\citealt*{goedhart03, goedhart04};
\citealt{goedhart09, goedhart14}; \citealt{araya10}; \citealt{szymczak11}; \citealt{fujisawa14}). They are characterized either 
by flaring or sinusoidal patterns with periods ranging from 29.5 to 668\,d and the relative amplitude from 0.2 to more than 30
\citep{fujisawa14}. Some of the sources show periodic behaviour in part of the spectrum, whereas others display synchronised 
variations of all maser features, sometimes with delays in flaring between individual features 
(\citealt{szymczak11}; \citealt{goedhart14}). The 12.2\,GHz maser emission, when observed, commonly shows periodic  
amplitude variations higher than those at 6.7\,GHz \citep{goedhart14}. G9.62+0.20E, an archetype of the class, exhibits a
regular flaring behaviour also in the 107\,GHz methanol maser line \citep{vanderwalt09}. Correlated periodic variability between 
the 6.7\,GHz methanol and 4.8\,GHz formaldehyde maser lines 
has been reported for G37.55$-$0.20 \citep{araya10}.

Different models have been proposed to explain the regularity of the methanol maser flares. For sources with periods of 
240-400\,d it has been postulated that a binary interaction between a HMYSO and its companion could periodically alter either 
the seed photon flux or pump rate in the maser region (\citealt{vanderwalt09, vanderwalt11}). Specifically, a colliding-wind binary
(CWB) could produce a periodic pulse in the ionising radiation that changes the electron density and periodically modulates 
the free-free emission from the \hii\, region against which the maser clouds are projected. 
In another model, the periodic maser variation is related to the flux of infrared emission that could be regulated by 
periodic accretion of material from the circumbinary disc onto a protostar or accretion disc \citep{araya10}.
Recently, \cite{parfenov14} proposed that the dust temperature variations in the circumbinary accretion disc are caused
by rotation of hot and dense material of the spiral shock wave in the disc central gap. The maser brightness depends on 
the density of the shocked material along the line between the central star and the maser region.
Simulations of stellar evolution indicate that a HMYSO still accreting at a rate larger than $10^{-3}M_{\sun}$yr$^{-1}$ 
and with a radius exceeding 100$R_{\sun}$ at maximum is pulsationally unstable \citep{inayoshi13}. The expected variations of 
luminosity with typical periods of 10$-$100\,d could drive the periodic variability of methanol masers 
with a sinusoidal light curve. 

\cite{goedhart05} mapped G9.62+0.20E at 12.2\,GHz using Very Long Baseline Interferometry (VLBI) and found that the brightness of the maser components increased 
during the course of a flare, whereas the overall structure was unchanged. The morphology of the 6.7\,GHz maser emission of the periodic 
source G22.357+0.066 was also unchanged in both observations near the burst maximum \citep{szymczak11}. These findings 
imply that the causes of the flares originate outside the maser region.

In this paper we report the discovery of periodic variations in four 6.7\,GHz methanol sources and produce an updated light curve of 
the known periodic source G22.357+0.066 (\citealt{szymczak11}) obtained as part of a long-term monitoring programme of a large sample of 
objects detected with the Torun 32\,m telescope (\citealt{szymczak12}). 

\section{Observations and data analysis}
134 target sources were selected from the Torun methanol maser catalogue (\citealt{szymczak12}) using coarse criteria intended 
primarily to monitor the objects with peak flux density greater than 5\,Jy and declination north of $-$11\degr. 
These objects stay above the elevation limit for relatively long durations and are suitable for a survey with 
the Torun 32\,m radio telescope. A detailed description of the sample will be presented in a subsequent paper. 
 
Each object was observed at least once a month during the period from 2009 June to 2010 October;  
then the objects showing a strong variability were observed at irregular intervals of 5$-$10\,d, until 2014 April.
Several gaps of 3$-$4 weeks in the monitoring campaign occurred due to scheduling constraints.

The 6668.519\,MHz methanol maser transition was observed in a frequency-switching mode using a dual-channel HEMT receiver mounted 
on the 32\,m radio telescope. The details of the observing setup have been described by \cite{szymczak14}. 
In summary, the system equivalent flux density was about 300\,Jy, the half-power beam width 5.5 arcmin and the pointing accuracy 
22\,arcsec. The autocorrelator was configured to record 4096 channels over 4\,MHz for each polarisation, yielding 
a spectral resolution of 0.09\,\kms~ after Hanning smoothing. The typical rms noise level in the spectra was 0.20--0.25\,Jy after 
averaging both polarisations. The system was calibrated by measuring the continuum point source 3C123 \citep{ott94}. 
Further checks were undertaken using the methanol maser source G32.745$-$0.076, in which some spectral features have not shown
variability higher than $\sim$8\,per cent (\citealt{szymczak14}). 
The errors on the absolute flux density measurements, estimated from the standard deviation of the flux density values of 
'non-variable' features of this source, were usually less than 10\,per cent.

\begin{figure}   
\resizebox{\hsize}{!}{\includegraphics[angle=0]{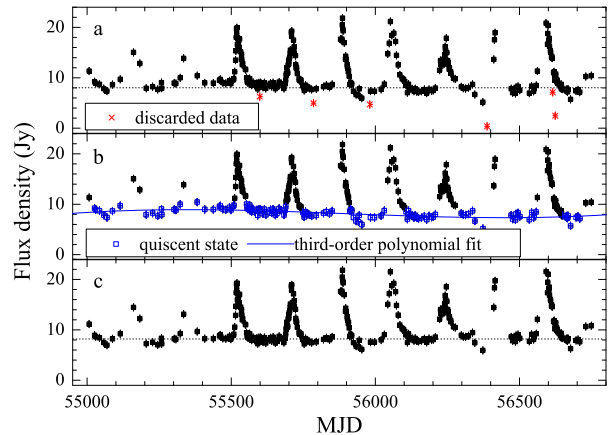}}
\caption{Effect of flagging and de-trending of the data on an example light curve. The light curve of 80.09\,\kms\, feature of source G22.357+0.066
is shown before flagging (a), after flagging but before de-trending (b) and after flagging and de-trending (c). 
The dotted horizontal line in panels a) and c) is put to mark the difference in the quiescent levels of the original and final curves.
\label{proc-effect}}
\end{figure}

The edited and flagged light curves were typically de-trended using a third-order polynomial or lower prior to period search.
Figure \ref{proc-effect} shows the effect of these procedures for the light curve of 80.09\,\kms\, feature of source G22.357+0.066.
Data points were discarded if their flux densities differed by $\ge5\sigma$ from the values observed just on the previous 
and the next day. This criterion was applied only for the intervals of quiescent emission. 
In the flare intervals sampled daily the individual observations were discarded only if the flux density obviously dropped 
below $\ge5\sigma$ as compared to the adjacent points. In effect about 2 per cent of observations of G22.357+0.066 were discarded
while for the rest sources the percentage of flagged observations is smaller. 
Slow and low amplitude changes of the flux density in the quiescent state were fitted with a third-order polynomial (Fig. \ref{proc-effect}).
In several cases a fit with a linear function was enough. 

In order to calculate the period and assess its significance the LS periodogram (\citealt{lomb76}; \citealt{scargle82}) was used. This well
established method is effective for unevenly spaced data as ours.  
For each periodically variable feature we obtained a normalised phase-folded flare profile by dividing the time-series into cycles
based on its best-fit period and the flux density values in each cycle were divided by the maximum observed in the cycle. 
We fitted each normalized light curve with a power function of the form: $S(t)=A^{s(t)} +C$, where $A$ and $C$ are constants
and $s(t)=[bcos(\omega t+\phi)/(1 - f sin(\omega t + \phi))]+a$. Here, $b$ is the amplitude measured relative to the average value $a$, 
$\omega=2\pi/P$,  where $P$ the period, $\phi$ the phase, and $f$ the asymmetry parameter defined as the rise time from the minimum to 
the maximum flux, divided by the period \citep{szymczak11}.
Time delays in flaring between features were calculated with the discrete correlation function \citep{edelson88}.  

\section{Results}
\begin{figure}   
\resizebox{\hsize}{!}{\includegraphics[angle=0]{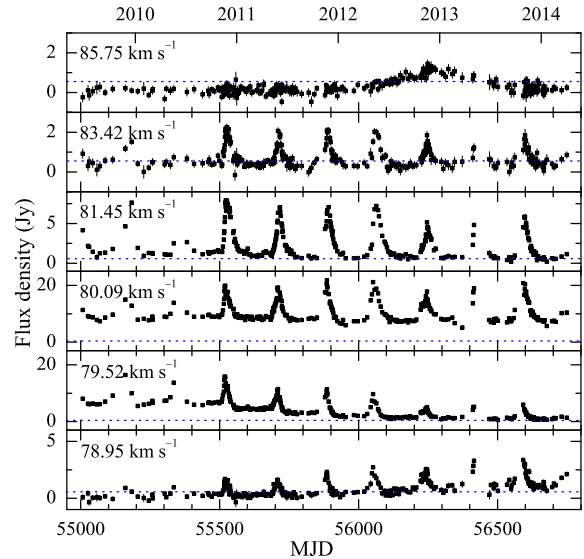}}
\caption{Light curves of selected 6.7\,GHz methanol maser features in G22.357+0.066. The dotted line in each panel marks the $3\sigma$ level. 
\label{lc_g22p357}}
\end{figure}

\begin{figure}   
\resizebox{\hsize}{!}{\includegraphics[angle=0]{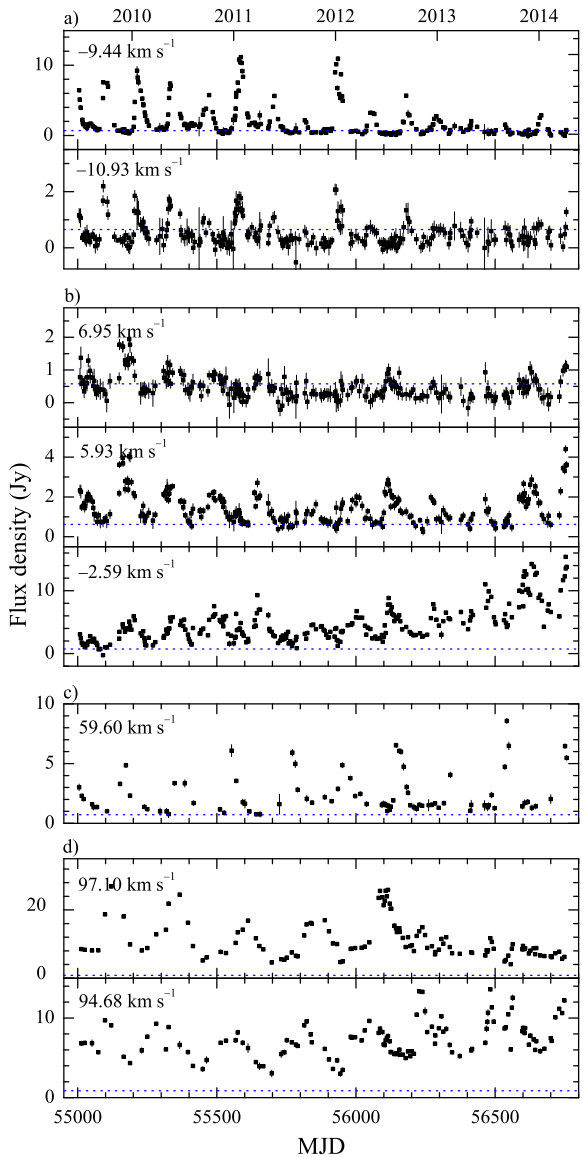}}
\caption{Light curves of flaring 6.7\,GHz methanol maser features in sources a) G75.76+0.34,
b) G73.06+1.80, c) G45.473+0.134 and d) G25.411+0.105. The dotted line in each panel marks the $3\sigma$ level. 
\label{lc_four-sources}}
\end{figure}

We have detected periodicity in five sources of which four are reported for the first time. Their average 6.7\,GHz spectra are shown in 
Figure \ref{period-spectr} and the plots of the flux density as a function of time and velocity are presented in Appendix A 
(Figs. \ref{ds-g22.357}--\ref{ds-g75.76}). The light curves of features with periodic behaviour are shown in Figures \ref{lc_g22p357} 
and \ref{lc_four-sources} and their parameters are given in Table \ref{periodic-features}. 
Because of the non-Gaussian nature of many of the spectral features their flux densities were read off from the spectra. 
We list the velocity of the feature,
the mean flux density, the relative amplitude, the period, the time-scale of variability, the ratio of the rise time to the decay
time ($\mathrm{R_{rd}}$), and the time delay of the flare between individual features. The relative amplitude is equal to 
(S$_\mathrm{max}$ $-$ S$_\mathrm{min}$)/S$_\mathrm{min}$, where S$_\mathrm{max}$ and S$_\mathrm{min}$ are maximum and minimum of 
the flux density, respectively. The time-scale of variability is the full width at half-maximum (FWHM) of the flare.

\begin{figure}   
\resizebox{\hsize}{!}{\includegraphics[angle=0]{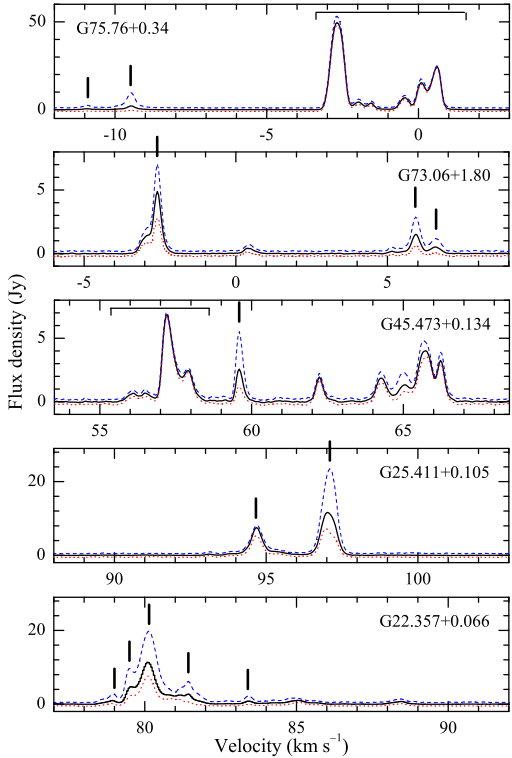}}
\caption{Spectra of 6.7\,GHz periodic methanol masers 
obtained at the positions given in Table \ref{periodic-features}. 
The solid (black) lines show the average spectra 
for the whole monitoring observations, while the dashed (blue) and dotted (red) lines show the average spectra
for intervals of high and low emission levels, respectively. The flaring features are marked by 
vertical bars. 
The velocity intervals of the emission from nearby but significantly ($\lesssim$78") offset sources are denoted by brackets.
\label{period-spectr}}
\end{figure}

\subsection{G22.357+0.066}
In the velocity range from 78.95 to 83.5\,\kms\, the emission showed periodic variations. Five permanent features 
with a mean flux density of 0.8$-$12\,Jy and relative amplitude of 2$-$7 varied with a period of 177.4$-$179.3\,d
(Fig. \ref{lc_g22p357}, Table \ref{periodic-features}). The mean period was 178.2$\pm$1.9\,d, where the error is a measure 
of the period dispersion for the features. A harmonic series is seen in the periodograms for all five features (Fig. \ref{periodograms}).   
The normalised phase-folded flare profiles are shown in Figure \ref{ph-cur1}. 
The ranges of their FWHM and $\mathrm{R_{rd}}$ are 28.6$-$32.7\,d and 0.68$-$1.83, respectively. 
The delays between the flare peaks of these features were 1$-$9\,d (Table \ref{periodic-features}).

Some features showed long-term changes of flux density in the quiescent state. For instance the 78.95\,\kms\, feature
initially was below the level of significant emission and then increased to 0.9\,Jy, whereas the feature 79.52\,\kms\, 
decreased from 16 to 0.8\,Jy over the monitoring period. 
For these features a systematic increase or decrease in the peak flux density occurred in  
consecutive cycles. In other periodic features the peak flux density commonly changed irregularly from one cycle to another.

The flux density of feature 83.42\,\kms\, in the quiescent state dropped below a typical $3\sigma$ sensitivity level of 0.6\,Jy.
We have made no attempt to estimate the intensity in this state with long integration observations, thus the relative amplitude (Table \ref{periodic-features}) 
gives a lower limit.
Faint emission with an intensity of 0.9$-$1.5\,Jy at maxima appeared after MJD 55885 at 78.64\,\kms\, and marginally detected emission 
(2.5$-$3.5$\sigma$) near 82.60 and 86.50\,\kms\, also showed flares synchronised with the flares of the main features 
(Fig. \ref{ds-g22.357}). No periodic variations were seen for the features near 85.00 and 88.43\,\kms. 

\begin{figure}   
\resizebox{\hsize}{!}{\includegraphics[angle=-90]{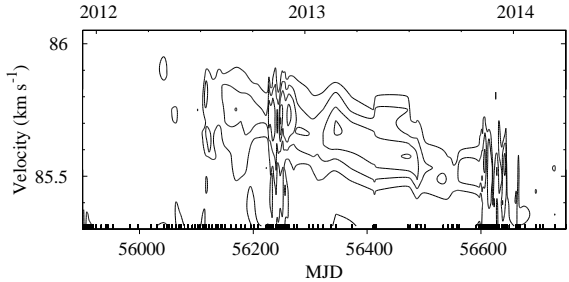}}
\caption{Maser burst in G22.357+0.066. The contours are at 2, 3, 4, and 5 times the 1$\sigma$ noise level of 0.26Jy.
The vertical bars in the bottom axis correspond to the dates of the observed spectra.
\label{g22.357-burst}}
\end{figure}

After MJD 56085 new emission appeared near 85.75\,\kms\, reaching a maximum of $\sim$1.3\,Jy near MJD 56248 (Fig. \ref{lc_g22p357}).
This burst lasted about 520\,d with no evidence of flux density modulation synchronised with the flaring activity of 
the permanent features. 
The line-width of the feature was 0.26$-$0.32\,\kms\, and uncorrelated with the flux density.  
The peak velocity showed a drift of 0.24\,\kms\,yr$^{-1}$ (Fig. \ref{g22.357-burst}). The event characteristics 
are similar to those reported in Cep\,A (\citealt{szymczak14}) but the signal-to-noise ratio in the spectra is less than 6 
and it is not clear whether the velocity drift is caused by variability of a close pair of spectrally blended features
(\citealt{szymczak14}) or by motion of the gas (\citealt{goddi11}).

\subsection{G25.411+0.105}
Periodic changes in the flux density of both spectral features occurred until MJD 56208, then the red-shifted emission at
97.10\,\kms\, slightly decreased without any periodic modulation, whereas periodic changes of the blue-shifted emission 
centred at 94.78\,\kms\, were seen throughout our monitoring (Figs. \ref{lc_four-sources} and \ref{ds-g25.411}). 
In the MJD interval of 55013--56208 the period was 245\,d (Table \ref{periodic-features}). 
The light curve of each feature has a sinusoidal-like shape with $\mathrm{R_{rd}}$ of 0.9$-$1.0 but the FWHM of the flare at
94.68\,\kms\, is approximately double that of the 97.10\,\kms\, (Fig. \ref{ph-cur2}, Table \ref{periodic-features}).

\subsection{G45.473+0.134}
The velocity range of the maser emission in this source is 11.2\,\kms\, (Fig. \ref{period-spectr}). The high resolution
maps revealed that the emission in the velocity range 
55.80$-$58.28\,\kms\, comes from the source G45.493+0.126 \citep*{bartkiewicz14} that is within the 32m antenna beam when pointing 
towards G45.473+0.134. A  clear periodicity of 195.7\,d is seen for the feature with the peak at 59.60\,\kms\, (Figs. \ref{lc_four-sources},
\ref{ds-g45.473} and \ref{periodograms}, Table \ref{periodic-features}). 
The light curve of this feature is characterised by irregular changes of the peak flare
from 3.6 to 8.6\,Jy, and a very stable flux density of 1.3$-$1.4\,Jy in the quiescent state. The FWHM is 34\,d and $\mathrm{R_{rd}}$=0.46
(Fig. \ref{ph-cur2}, Table \ref{periodic-features}). The features in the velocity range from 64.2 to 66.3\,\kms\, show 
weak flares synchronised with the flares of the 59.60\,\kms\, feature but only in some cycles. The emission at 62.23\,\kms\, 
is constant within less than 10\,per cent over the whole monitoring period.

\subsection{G73.06+1.80}
Features at $-$2.59 and 5.93\,\kms\, showed variations with a period of 160\,d (Figs. \ref{lc_four-sources}, \ref{ds-g73.06} and
\ref{periodograms}, Table \ref{periodic-features}). The weak emission at 6.95\,\kms\, flared synchronously with the strongest features 
in some cycles only. The flare peak flux density varied significantly from cycle to cycle for all features. 
In the quiescent state the emission of feature $-$2.59\,\kms\, increased from 0.5 to 6\,Jy, whereas that
of feature 5.93\,\kms\, was constant within $3\sigma$.
The light curves of the periodic features were strongly asymmetric (Fig. \ref{ph-cur2}) with $\mathrm{R_{rd}}$ 
of 0.27$-$0.35 and the time-scale of variability of 60$-$66\,d (Table \ref{periodic-features}).  
The mean delay between the flaring features was $13.8\pm3.6$\,d.

\subsection{G75.76+0.34}
The maser emission at velocities lower than $-$9\,\kms\, exhibited a flaring behaviour with a period of 119.9\,d.
The intensity of flare peaks was clearly modulated at least over the first 12 cycles since MJD 55158 and 
the periodograms for features $-$10.93 and $-$9.44\,\kms\, revealed less significant (p$<$0.005) periods of 348.8 and 328.6\,d, 
respectively (Figs. \ref{lc_four-sources}, \ref{ds-g75.76} and \ref{periodograms}). The flares were asymmetric with $\mathrm{R_{rd}}$ of
0.63$-$0.72 and their FWHM were about 33\,d (Fig. \ref{lc_four-sources}, Table \ref{periodic-features}). 
The emission of feature $-$9.44\,\kms\,, in some quiescent intervals, dropped below a typical $3\sigma$ sensitivity level  
of 0.7\,Jy, so that the relative amplitude of 7.6 (Table \ref{periodic-features}) gives the lower limit. The flux density of feature $-$10.93\,\kms\, always dropped
below this threshold. No attempt was made to determine a low state flux density with long integration observations. 
The emission in the velocity range from $-$3.2 to 1.0\,\kms\, did not show periodic variability. 
The JVLA data\footnote{http://bessel.vlbi-astrometry.org/ (Hu, priv.comm.)} 
imply that this emission comes from the source G75.782+0.342 which is located $\sim$76" north-east from 
the maser source G75.761+0.340 showing periodic variations of both features $-$10.93 and $-$9.44\,\kms.

\begin{table*}
 \caption{The parameters of periodic maser features. 
\label{periodic-features}}
\begin{tabular}{c c c c c r r l c c}
\hline
Source   & RA(2000) & Dec(2000) & Feature       &Mean         &Relative  & Period  & Time-scale      &  R$_\mathrm{rd}$ & Time  \\
name     & (h     m      s) & ($^{\circ}$      '       ") & velocity      &flux density &amplitude &         & of variability &                  & delay  \\
         & & &(km\,s$^{-1}$) &   (Jy)      &          &(day)    & FWHM (day)     &                  & (d)    \\
\hline
G22.357+0.066 & 18 31 44.1 & $-$09 22 12.3 &78.95 & 0.92(0.04)  & 3.5(0.4) & 179.3 & 32.7(1.7) & 1.83(0.39) & 0.85(0.55)  \\
              &            &               & 79.52 & 4.56(0.20)  & 4.2(0.6) & 177.4 & 28.9(1.1) & 1.50(0.28) & 1.81(1.14)  \\
              &            &               & 80.09 & 11.19(0.22) & 2.0(0.1) & 179.3 & 29.8(1.0) & 0.68(0.08) & 0           \\
              &            &               & 81.45 & 2.60(0.12)  & 7.0(0.7) & 177.4 & 32.1(1.5) & 0.76(0.12) & -7.00(3.38) \\
              &            &               & 83.42 & 0.78(0.03)  & $>$4.2(0.3) & 177.4 & 28.6(1.6) & 1.08(0.27) & -4.75(2.59) \\
G25.411+0.105$^{\mathrm{a}}$  & 18 37 16.9  & $-$06 38 30.5 & 94.68 & 7.34(0.20)  & 1.2(0.1) & 244.7 & 121.7(2.1) & 1.00(0.18) & - \\
                              &             &               & 97.10 & 11.19(0.50) & 2.1(0.1) & 244.7 & 68.9(1.8) & 0.91(0.13) & - \\
G45.473+0.134 & 19 14 07.3 & 11 12 16.4   & 59.60 & 2.44(0.18)  & 3.5(0.6) & 195.6 & 34.0(1.2) & 0.46(0.10) & - \\
G73.06+1.80   & 20 08 10.1 & 35 59 24.7   &$-$2.59& 4.67(0.18)  & 1.9(0.6) & 159.0 & 59.5(2.4) & 0.27(0.18) &  13.8(3.6)$^{\mathrm{b}}$ \\
              &            &              & 5.93  & 1.45(0.05)  & 2.3(0.3) & 159.0 & 66.1(2.0) & 0.35(0.15) &   0 \\
G75.76+0.34   & 20 21 40.1 & 37 25 37.0   &$-$10.93&0.52(0.03) & $>$5.4(0.8)  & 119.9 & 33.1(1.9) & 0.63(0.14) & - \\
              &            &              &$-$9.44& 2.00(0.14) & $>$7.6(1.3)  & 119.9 & 32.5(1.3) & 0.72(0.11) & - \\
\hline              
\end{tabular}

\noindent
$^{\mathrm{a}}$ the parameters derived for MJD interval of 55013$-$56208, 
$^{\mathrm{b}}$ the mean for cycles 1-3, 6 and 11
\end{table*}

\section{Discussion}
\subsection{Period statistics}
A sample of 134 methanol masers with a peak flux density greater than 5\,Jy and declination north 
of $-$11\degr\, selected from the Torun methanol maser catalogue (\citealt{szymczak12}) was monitored with 
the 32\,m radio telescope. We found that about 4 per cent (5/134) of the sources show periodic variations with 
periods ranging from 120 to 245\,d. 
In order to examine the sensitivity of our observations to periodic sources we have injected periodic signals of different
levels and periods into real light curves and estimated the reliability with which they are appeared in the periodograms.
For the light curve of G22.357+0.066 consisting 279 observations a significant frequency (a false alarm probability of 0.01 per cent)
in the periodogram occurred for signals with the relative amplitude $\ge$0.26 and period from 20 to 500\,d.   
In the case of source G45.473+0.134 with only 86 observations a periodic injected signal of 30$-$500\,d was significant for
the relative amplitude $\ge$0.41. We conclude that our observations with average cadences of $\sim$6\,d (G22.357+0.066) and 
$\sim$21\,d (G45.473+0.134) are complete, in terms of ability of periodic maser detection, for the relative amplitudes 
higher than 0.26 and 0.41, respectively. 
Our detection rate of periodic masers is a factor of three smaller than those reported by \cite{goedhart04} who found 
7 periodic sources from their 54 targets.
This significant difference is likely due to a selection effect; Goedhart et al.'s sample basically relied
on a list of targets previously known as variables, while we studied a flux limited sample without prior
information on the variability.

\begin{figure}   
\resizebox{\hsize}{!}{\includegraphics[angle=0]{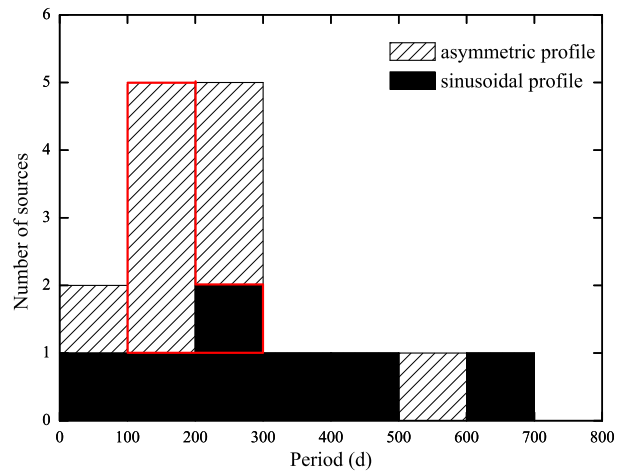}}
\caption{Histogram of periods for periodic 6.7\,GHz methanol maser sources. The objects with asymmetric flare profile 
and sinusoidal-like profile are shown in blue and red, respectively. The contribution from the Torun 32m telescope
observations are marked by the red line. 
\label{histogram-period}}
\end{figure}

Figure \ref{histogram-period} shows the period distribution for the 16 periodic methanol sources known to date
as compiled by \citet{fujisawa14} (their table 2) supplemented by our new results and refined data from \citet{goedhart14}. 
The period ranges from 29.5 to 668\,d with a median value of 210\,d. Our detections have a rather narrow range 
of period from 120 to 245\,d. 
The period distribution may suggest that the intrinsic fraction of periodic masers decreases with the period length. 

\begin{figure}   
\resizebox{\hsize}{!}{\includegraphics[angle=0]{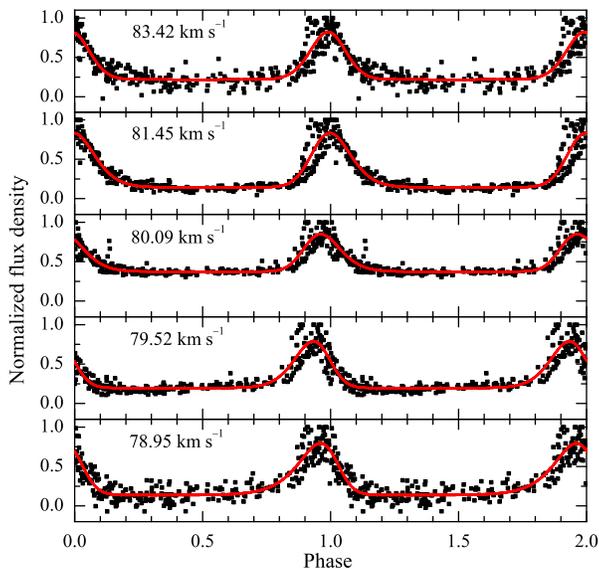}}
\caption{Phase-folded normalized light curves of selected features of G22.356+0.066. Each panel is labelled with the peak velocity.
\label{ph-cur1}}
\end{figure}

\subsection{Maser flare profile}
We note that 9 out of 16 periodic masers are objects where the representative feature has an asymmetric flare profile, i.e.
with a fast rise and relatively slow fall of intensity. These are the following sources G9.62+0.20E (\citealt{goedhart03, goedhart14}), 
G22.357+0.066, G45.473+0.134, G73.06+1.80, G75.76+0.34 (Table \ref{periodic-features}), G37.55+00.20 (\citealt{araya10}),
G107.29+5.64 (\citealt{fujisawa14}), G328.24-0.55 and G331.13-0.24 (\citealt{goedhart14}).
The asymmetric flare profile can be readily explained with the simple CWB model (\citealt{vanderwalt11}).
In this model a periodic pulse of ionizing radiation enhances the electron density in a volume of partially ionized
gas against which the maser clouds are projected. The intensity of maser emission varies synchronously with the periodic
variations of the background radiation amplified by the maser process. The different shapes and amplitudes of the maser flares 
in individual sources can be satisfactorily reproduced by modifying the orbital parameters of the binary system 
(\citealt{vanderwalt11}; \citealt{goedhart14}).

\begin{figure}   
\resizebox{\hsize}{!}{\includegraphics[angle=0]{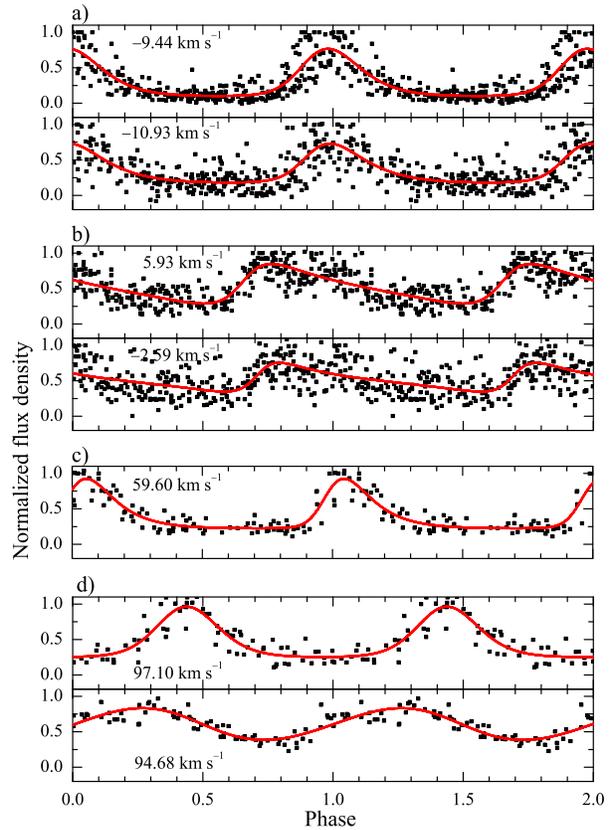}}
\caption{Phase-folded normalized light curves of selected features of four targets;
a) G75.76+0.34, b) G73.06+1.80, c) G45.473+0.134 and d) G25.411+0.105. Each panel is labelled by the velocity of feature. 
\label{ph-cur2}}
\end{figure}

The mean profiles of the flares in G45.473+0.134, G73.06+1.80 and G75.76+0.34 (Fig. \ref{ph-cur2}) are generally consistent 
with the CWB model. The changes in flare profiles between cycles and features of the two last objects 
(Fig. \ref{lc_four-sources}) may suggest that the seed photon flux is not constant and the maser optical depth
shows long-term changes. The variability patterns of the features at 80.09 and 81.45\,km\,s$^{-1}$ with 
R$_{\mathrm{rd}}\approx$0.72 of G22.357+0.066 are similar to those reported for G9.62+0.20E (\citealt{goedhart09}) 
but they are quite different from those of the features 78.95 and 79.52\,\kms\, with R$_{\mathrm{rd}}\ge$1.50 (Fig. \ref{ph-cur1}). 
Such widely different patterns of the features cannot be explained by changes in the background free-free emission 
postulated in the simple CWB model (\citealt{vanderwalt11}). An alternative explanation proposes a model of the circumbinary 
accretion disc (\citealt{parfenov14}) in which the pattern of the maser flare follows that of the gas column density 
along the line between the central star and the maser region. For the specific structure of the disc (\citealt{sytov09}), 
the column density changes with the inclination angle of the binary system, so the flare profile with R$_{\mathrm{rd}}<$1.0 
may be generated in the molecular disc seen edge-on ($i$=90\degr). A small change of disc inclination on the line
of sight ($i$=83\degr) would result in a flare profile with R$_{\mathrm{rd}}>$1.0. This issue needs detailed calculations
and is beyond the scope of the paper. 

G25.411+0.105 has significantly different light curves for the two main features (Fig. \ref{lc_four-sources}). 
The maximum of the 94.68\,\kms\, feature leads in phase by $\sim0.15P$ relative that of 97.10\,\kms\, feature while
both features have a minimum at the same phase.
Similar characteristics were reported for G12.89+0.49 and
G331.13-0.24 (\citealt{goedhart14}) and could be the result of an amplification of the background emission from a processing outflow
or thermal jet both of which are not directly associated with the methanol maser clouds (\citealt{macleod96}).
\cite{hofner11} detected in G25.411+0.105 a weak compact continuum source at 25.5\,GHz with a spectral index of 0.6 between 19.3 and 25.5\,GHz
at the western edge of the maser distribution (Fig. \ref{periodic-evn-three}). The observed parameters of 
this continuum source imply a brightness temperature of $\sim$5400\,K. For a linear maser geometry (\citealt{watson03})
in unsaturated regime with $R/\Gamma<0.2$, where $R$ is the pump rate and $\Gamma$ is the decay rate, a modulation of 
the background emission can produce the maser flux density variations (Sobolev \& Parfenov, priv. comm.). 
The disappearance of cyclic changes and weak decay of the 97.10\,\kms\, feature after MJD 56208 resemble those reported 
for G188.95+0.89 (\citealt{goedhart09}) which has been interpreted in terms of the recombination of the ionized gas
against which the maser cloud is projected.

\subsection{Size of maser region showing periodic variability}
\begin{figure}   
\centering
\includegraphics[angle=0, width=0.8\columnwidth,trim=0.05cm 0.05cm 0.35cm 0.35cm,clip]{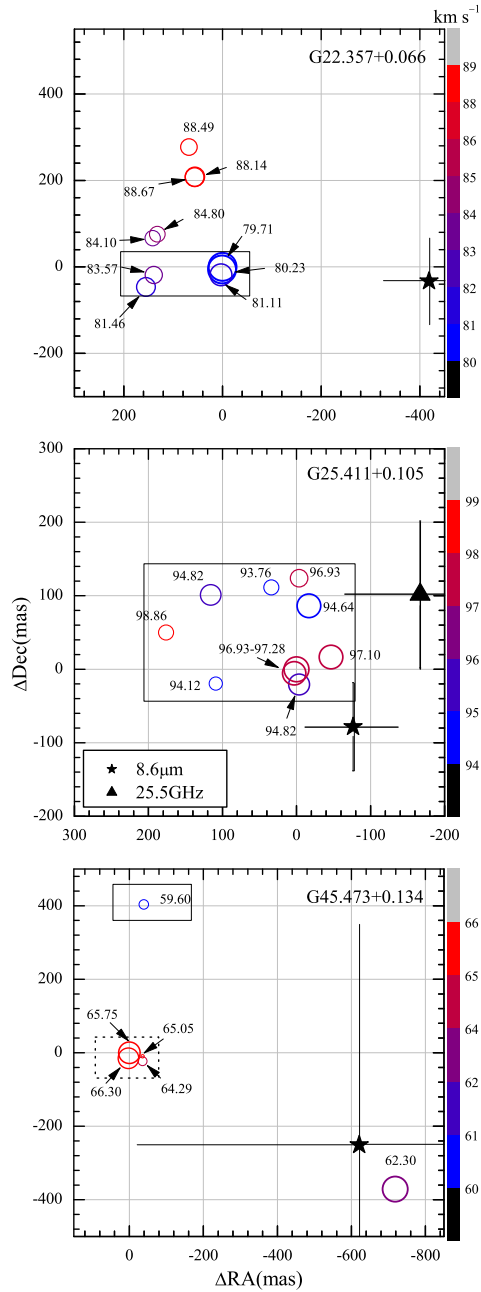}
\caption{Maps of the 6.7\,GHz methanol maser in three periodic sources (\citealt{bartkiewicz09, bartkiewicz14}).
The circle size is proportional to the logarithm of maser brightness of cloud and its colour corresponds
to the velocity. The individual maser clouds are also labelled by the velocity in km\,s$^{-1}$.
The groups of maser clouds showing clear and marginal periodic variations are framed by 
the solid and dotted rectangles, respectively.
The star symbol with error bars in each panel indicates the location of the nearest infrared source for 
G22.356+0.066 and G45.473+0.134 (\citealt{lucas08}) and G25.411+0.105 (\citealt{debuizer12}).
The triangle symbol with error bars shows the position of the radio continuum source (\citealt{hofner11}). 
\label{periodic-evn-three}}
\end{figure}

G22.357+0.066, G25.411+0.105 and G45.473+0.134 have been observed at 6.7\,GHz with milliarcsecond (mas) 
resolution (\citealt{bartkiewicz09, bartkiewicz14}). Their maser cloud maps are shown in Figure \ref{periodic-evn-three} 
together with the position of the nearest infrared source for which the spectral density distribution is consistent with an HMYSO candidate.
The groups of the maser clouds with clear and marginal periodic variability are marked. 

In G22.357+0.066, the clouds showing flares are located in an elongated area of size $<$156$\times$29\,mas 
(Fig. \ref{periodic-evn-three}) that corresponds to a projected linear size of $<760\times140$\,au for the near kinematic 
distance of 4.86\,kpc (\citealt{reid09}). The maser clouds without or with marginal variations of the intensity are 
in the northern edge of the distribution of size $>$1500\,au. VLBI observations at two epochs spanning two years have shown that 
the spatial structure of the source on a mas scale remains unchanged (\citealt{szymczak11}). 
The maser spectra obtained with the VLBI and single dish at nearly the same epoch are very similar. This means that
almost all maser clouds have compact cores and the source structure was fully recovered with a 6\,mas$\times$12\,mas beam (\citealt{bartkiewicz09}).
There is a 5\,d delay in the peak of the flares between the 80.09 and 83.42\,\kms\, features (Table \ref{periodic-features}) 
that can be explained by light travel time between the maser clouds of 80.23 and 83.57\,\kms\, (Fig. \ref{periodic-evn-three}) 
that yields a distance of 850\,au between the regions. This agrees within the error range with the projected distance between 
these clouds of 700\,au. Thus, the geometry of maser clouds and time delay measurements clearly indicate that an exciting source 
is located in the west side of the maser distribution, consistent with the location of the infrared HMSYO candidate marked 
in Figure \ref{periodic-evn-three}. The brightness temperature ($T_{\mathrm b}$) of the maser clouds in G22.357+0.066 derived from Bartkiewicz et al.'s 
data ranges from $6\times 10^7$ to $4\times 10^9$\,K. The mean values of $T_{\mathrm b}$ for the clouds with periodic and 
non-periodic emission are $1.7\pm0.9\times 10^9$ and $1.3\pm0.3\times 10^8$\,K, respectively. A cluster of the brightest maser 
clouds showing periodic variations is located closer to the exciting source than the weaker clouds with or without periodic behaviour.

The above summarised observational characteristics of G22.357+0.066 do not support the hypothesis that the maser flares are related to changes 
in the background free-free emission. The maser region with flaring behaviour has a size more than one order of magnitude smaller 
than the typical size ($\le$10000\,au) of hyper-compact \hii\, regions (\citealt{hoare07}). 
No obvious mechanism exists to create such small scale fluctuations in the background radiation.
Moreover, as the clouds showing periodic variations are brighter than those without or with marginal variations one can suppose that
the ratio of $R/\Gamma$ is higher in the first group of clouds than in the second group. Thus, the flux density of flaring components 
is expected to be less dependent on the variations of the background photon flux. 

The observations of G22.357+0.066 can support the hypothesis that the maser flares are related to changes in the maser optical depths
and excitation temperature in some maser regions periodically triggered by a pulse from the massive exciting star or binary system.
Periodic pulses of UV radiation are predicted in the model of a pulsationally unstable protostar (\citealt{inayoshi13}) but it is
difficult to explain why only some parts of the maser regions in G22.357+0.066 are excited. The same difficulty  
applies to the scenario where periodic variations of the dust temperature are caused by variability in the accretion
rate in a binary system (\citealt{araya10}).
The illumination of the molecular disc by the bow shock hot and dense material of the spiral shock in the circumbinary disc central gap 
(\citealt{parfenov14}) could explain the periodic maser flare of some part of maser regions, in a particular condition when 
the maser emission arises in a disc seen edge-on. 

We conclude that any mechanism responsible for the periodic changes of the maser emission must be able 
to influence only part of the masing region if it is to be consistent with observations.

VLBI observations of G45.473+0.134 in March 2010 have shown that the maser emission comes from three clusters (\citealt{bartkiewicz14}). 
No significant changes in morphology were seen as compared with the maps obtained in 2007 with MERLIN (\citealt{pandian11}). 
Comparison of the spectra taken with the 32\,m telescope and VLBI indicates that less than 15 per cent of the flux density was missed
in the VLBI observation (\citealt{bartkiewicz14}, their figure A1), thus the source morphology is accurately recovered. 
The periodic flares originate from a single cloud at 59.60\,\kms\, (Fig. \ref{periodic-evn-three}) of $T_{\mathrm{b}}=7.5\times10^8$\,K. 
In the map taken with MERLIN (\citealt{pandian11}) the maser distribution has a double structure of size less than 70\,mas that corresponds 
to 480\,au for the distance of 6.9\,kpc (\citealt{pandian09}). 
The cloud showing the periodic emission is separated by 413\,mas (2850\,au) from the cluster of four clouds at velocities of 64.29$-$66.30\,\kms. 
This cluster shows weak periodic emission only during some cycles well synchronised with the flares of 59.60\,\kms\, emission.  
No periodic emission is seen from the 62.30\,\kms\, cloud that is located 1.03" (7100\,au) south-west from the site of the periodic emission. 
We conclude that the mechanism causing periodic changes of the flux density is effective only in a relatively small ($<$480\,au) region
of the source. As the position uncertainty of the candidate of HMYSO observed in the infrared (Fig. \ref{periodic-evn-three}) is comparable with
the size of the overall maser distribution it is plausible that it may drive the maser periodicity.

G25.411+0.105 shows a ring-like distribution of the maser clouds (Fig. \ref{periodic-evn-three}) with a diameter of 
190\,mas (\citealt{bartkiewicz09}) that corresponds to 1000\,au for the distance of 5.35\,kpc (\citealt{debuizer12}).  
The maser emission does not show any velocity gradient; some blue- and red-shifted clouds are considerably blended spatially and
in some spectral channels (e.g. 94.82\,\kms) the emission appears at very different parts of the ring. In such a case the light
curve of the blue-shifted emission can significantly differ from that of the red-shifted emission as we observed with the single dish.
The range of brightness temperatures of the maser components is $0.3-12\times10^8$\,K.
The unresolved (0.8"$\times \le$0.6``) continuum source at 25.5\,GHz (\citealt{hofner11}) coincides within 200\,mas with 
the strongest maser component. This continuum emission is interpreted as an ionized jet at the base of a massive flow observed 
in the CO lines (\citealt{hofner11}) and it may induce the maser variations. If the source could be detected at cm wavelengths
it would be possible to evaluate the effect of the background emission on the maser flaring.

We have inspected the first three objects listed in Table \ref{periodic-features} for continuum emission at 5\,GHz using 
the Cornish survey (\citealt{hoare12}; \citealt{purcell13}). No emission was seen at the $5\sigma$ level of 2\,mJy.
We found no evidence for cm radio emission in the remaining targets when searching the literature using the SIMBAD database.
One can speculate that our periodic sources are associated with HMYSO in binary systems in very early phases with
vigorous accretion processes.

\section{Conclusions}
We have found four new periodic methanol sources at 6.7\,GHz and updated the light curve of the previously
known source G22.357+0.066. This study increases the number of known periodic masers to 16 with a median
period of 210\,d. The period distribution hints to an association of the periodic masers with massive binary systems.
The diversity of variability patterns cannot be easily interpreted within the frame of a simple model of
the colliding-wind binary system which provides periodic variations of the continuum emission from the 
background \hii\, region. The study revealed that the periodic variations can be confined to a single feature in 
the spectrum for some sources. In two sources, the maser flares originate in a region of linear size
less than 480-760\,au that contributes 20-50 per cent of the overall maser distribution. This implies that
the mechanism generating periodic maser variations affects only some parts of the masing regions.

\section*{Acknowledgements}
We would like to thank the TfCA staff and the students for assistance with the observations. 
We are grateful to Eric G\'erard for careful reading and helpful comments on the manuscript 
and Andrej Sobolev and Sergey Parfenov for useful discussions. We thank Bo Hu for providing us
the data on source G75.76+0.34 before publication 
and the referee Simon Ellingsen for a very thorough reading of the manuscript.
This research has made use of the SIMBAD database, operated at CDS (Strasbourg, France) and
NASA's Astrophysics Data System Bibliographic Services. 
The work was supported by the Polish National Science Centre grant 2011/03/B/ST9/00627.

\begin{appendix}

\section{The dynamic spectra and L-S periodograms}

For completeness, in this appendix, we present the dynamic spectra of the sources and
the Lomb-Scargle (L-S) periodograms for the features for which significant frequencies were found.

\begin{figure}
\centering
\includegraphics[width=1.0\columnwidth,trim=-0.05cm -0.05cm -0.05cm -0.05cm,clip]{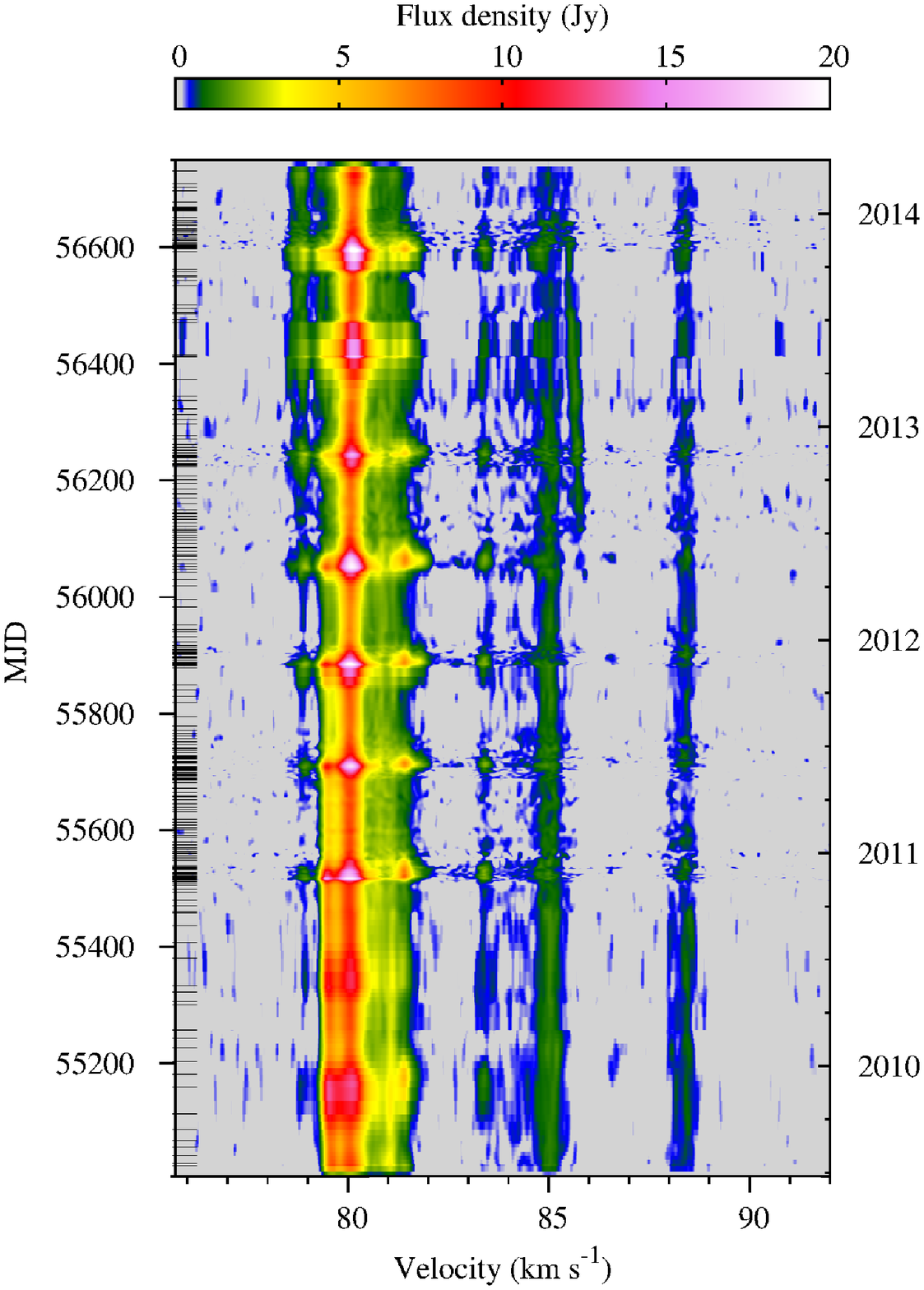}
\caption{False-colour image of the 6.7 GHz methanol maser flux density over velocity and time for 
G22.357+0.066. The velocity scale is relative to the local standard of rest. The horizontal
bars in the left coordinate correspond to the dates of the observed spectra.}
\label{ds-g22.357}
\end{figure}

\begin{figure}
\centering
\includegraphics[width=1.0\columnwidth,trim=-0.05cm -0.05cm -0.05cm -0.05cm,clip]{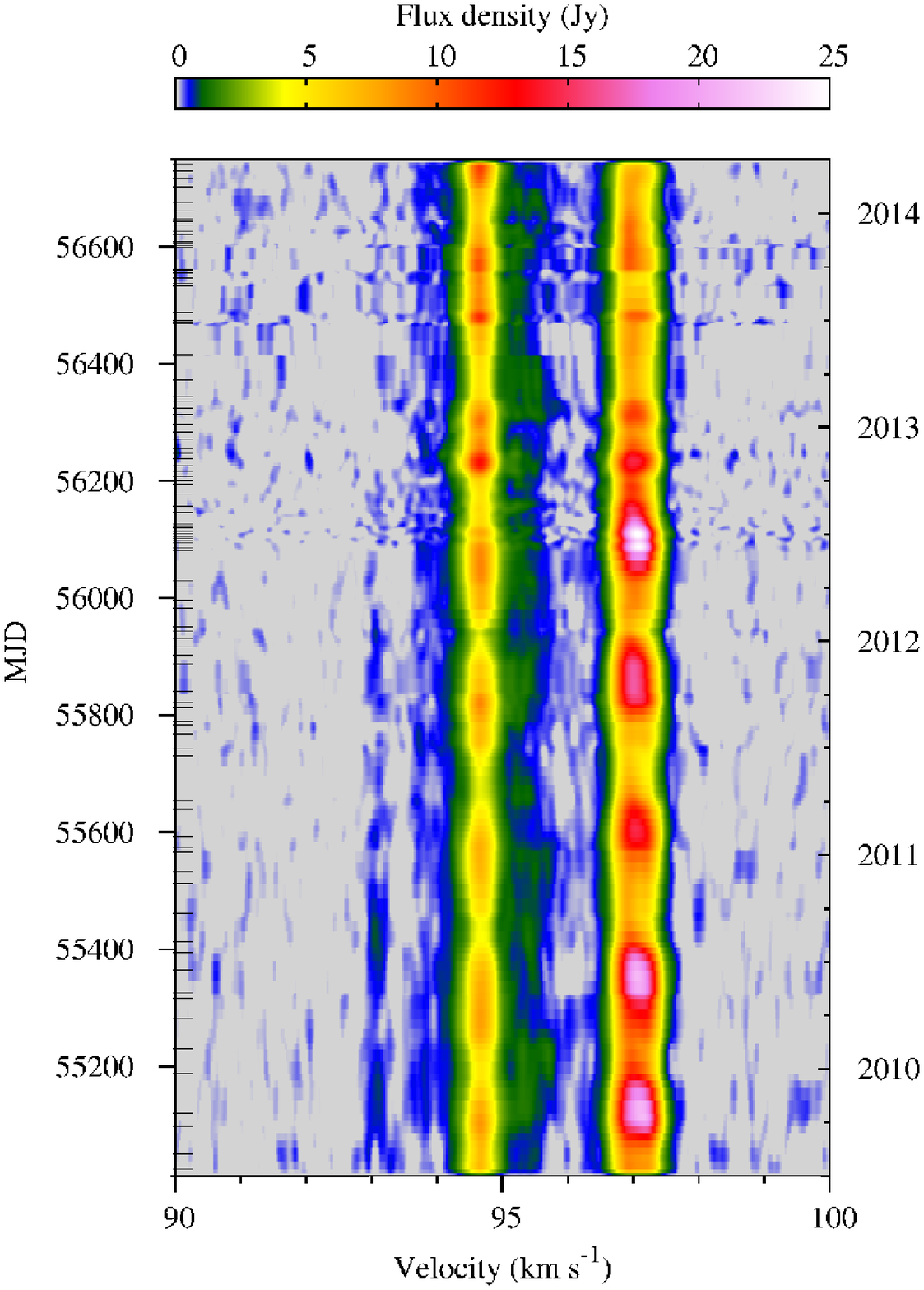}
\caption{Same as in Fig. \ref{ds-g22.357} but for source G25.411+0.105.}
\label{ds-g25.411}
\end{figure}

\begin{figure}
\centering
\includegraphics[width=1.0\columnwidth,trim=-0.05cm -0.05cm -0.05cm -0.05cm,clip]{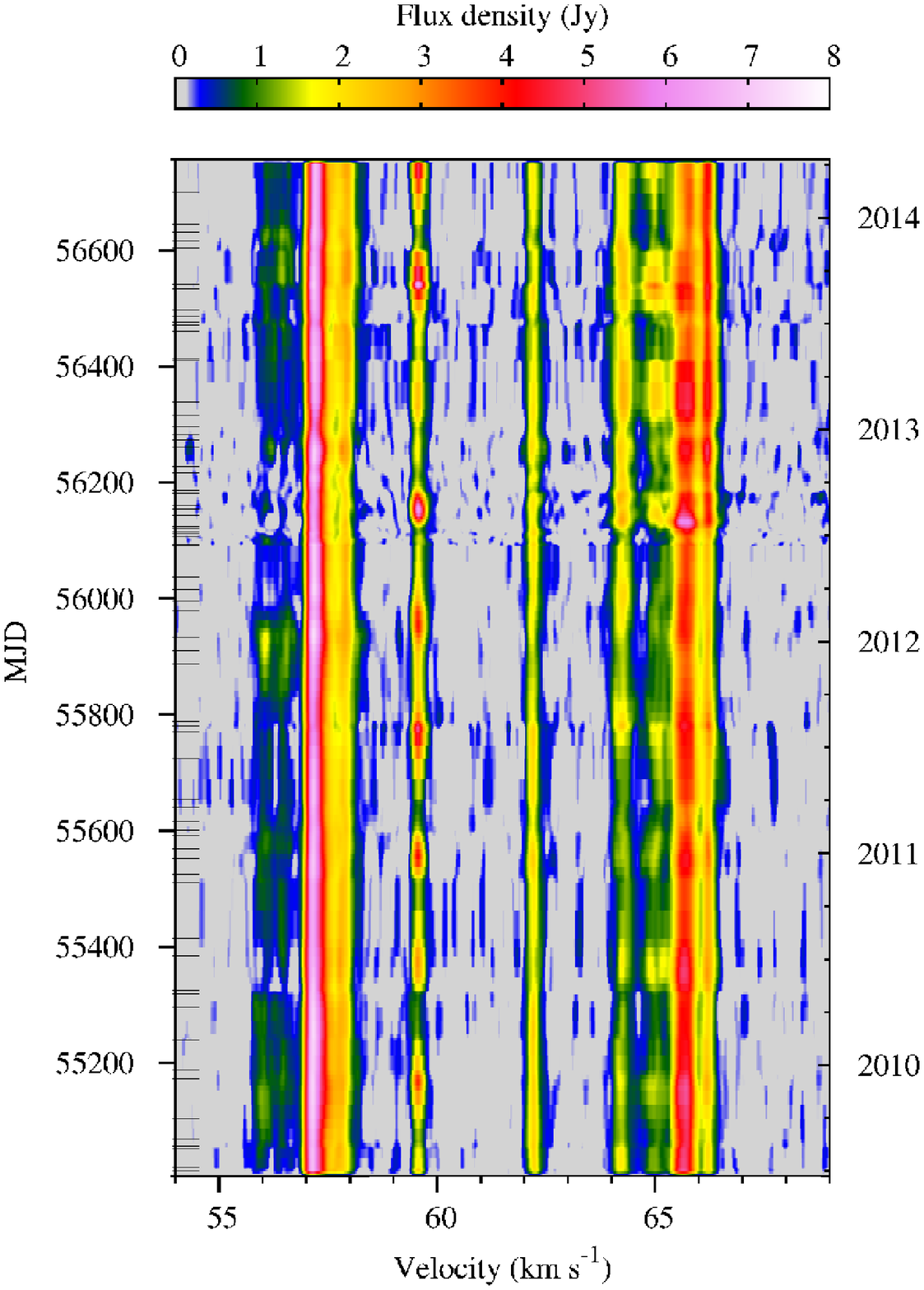}
\caption{Same as in Fig. \ref{ds-g22.357} but for source G45.473+0.134.}
\label{ds-g45.473}
\end{figure}

\begin{figure}
\centering
\includegraphics[width=1.0\columnwidth,trim=-0.05cm -0.05cm -0.05cm -0.05cm,clip]{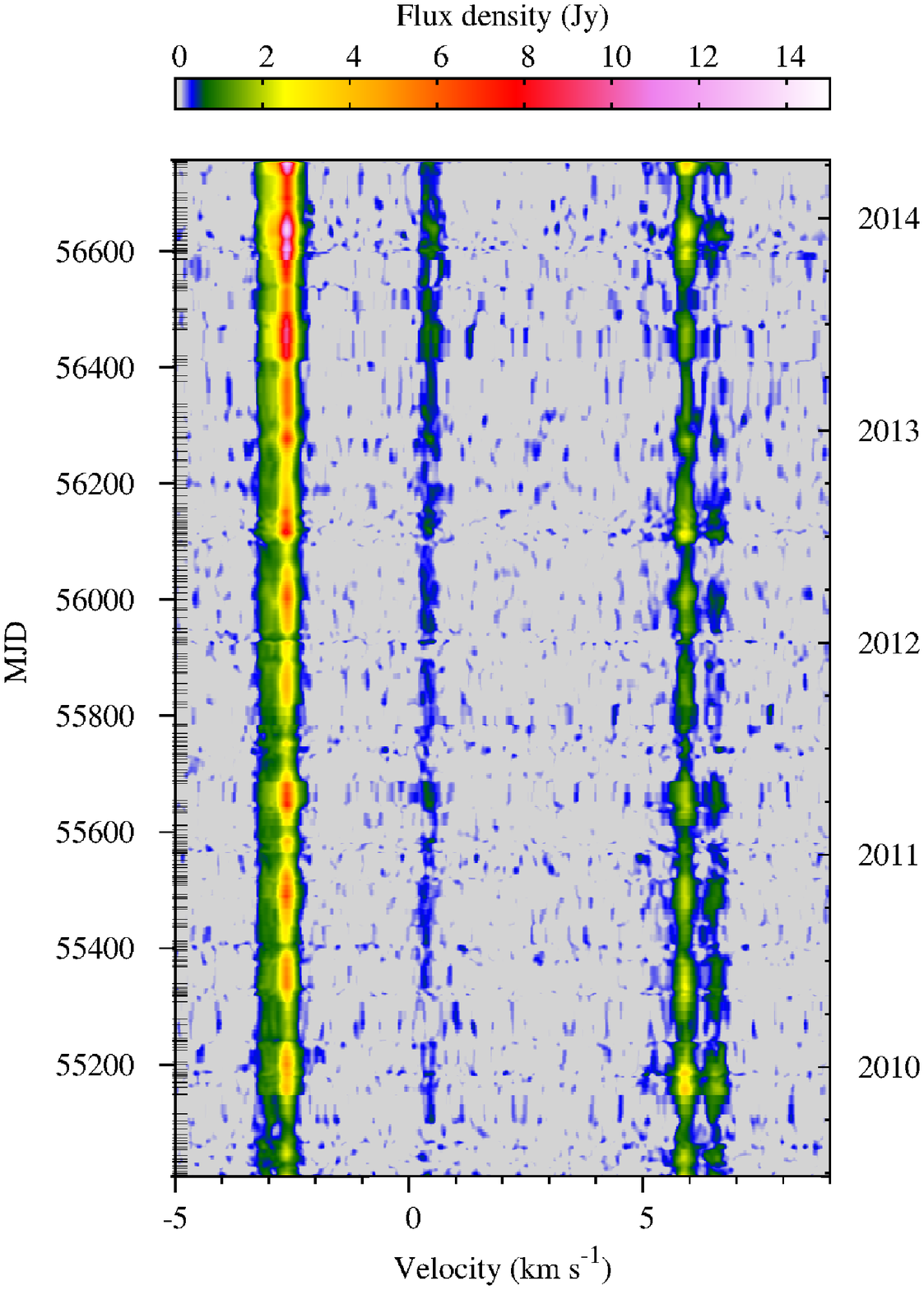}
\caption{Same as in Fig. \ref{ds-g22.357} but for source G73.06+1.80.}
\label{ds-g73.06}
\end{figure}

\begin{figure}
\centering
\includegraphics[width=1.0\columnwidth,trim=-0.05cm -0.05cm -0.05cm -0.05cm,clip]{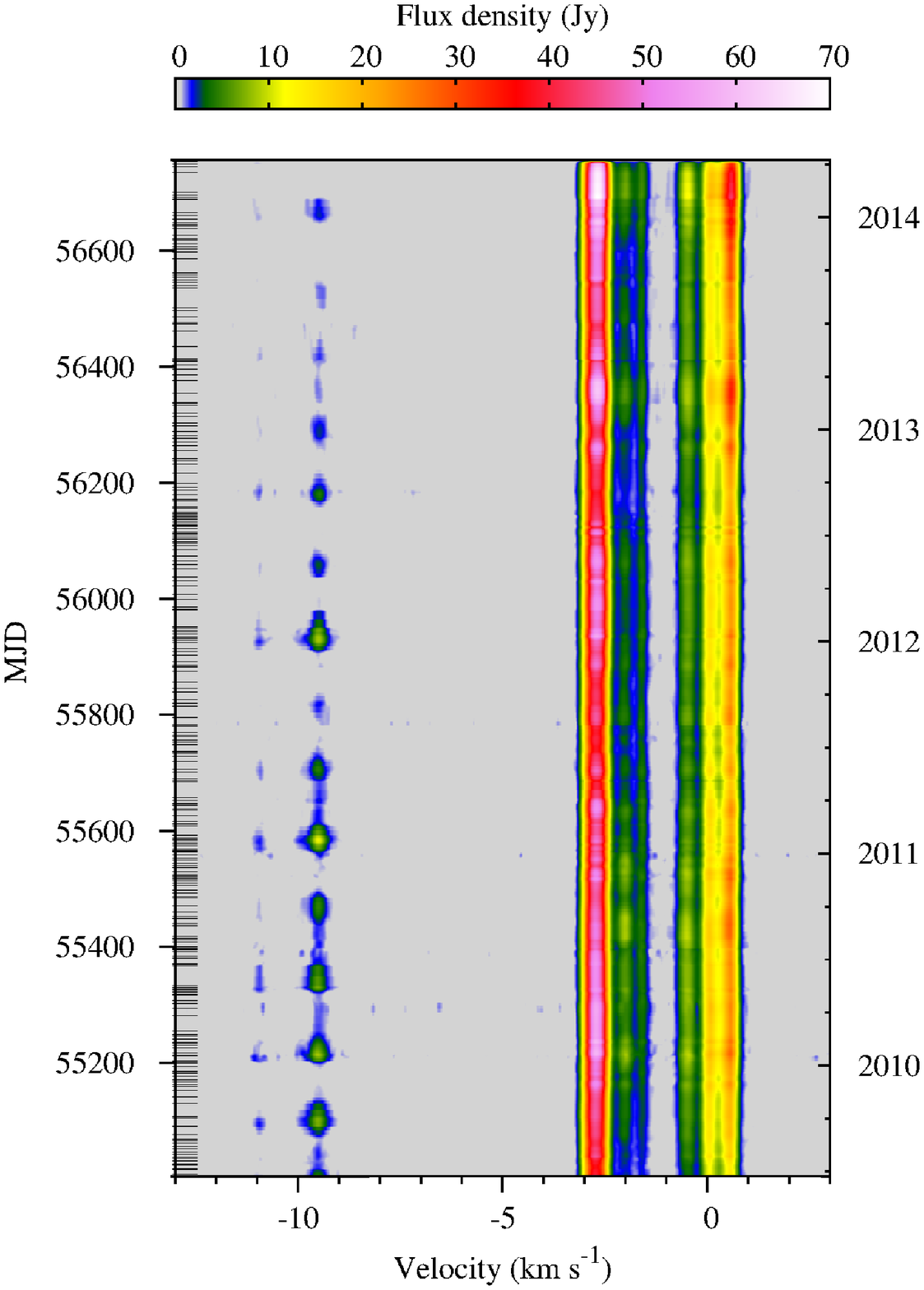}
\caption{Same as in Fig. \ref{ds-g22.357} but for source G75.76+0.34.}
\label{ds-g75.76}
\end{figure}

\begin{figure*}
\centering
\includegraphics[width=2.0\columnwidth,trim=-0.15cm -0.15cm -0.15cm -0.15cm,clip]{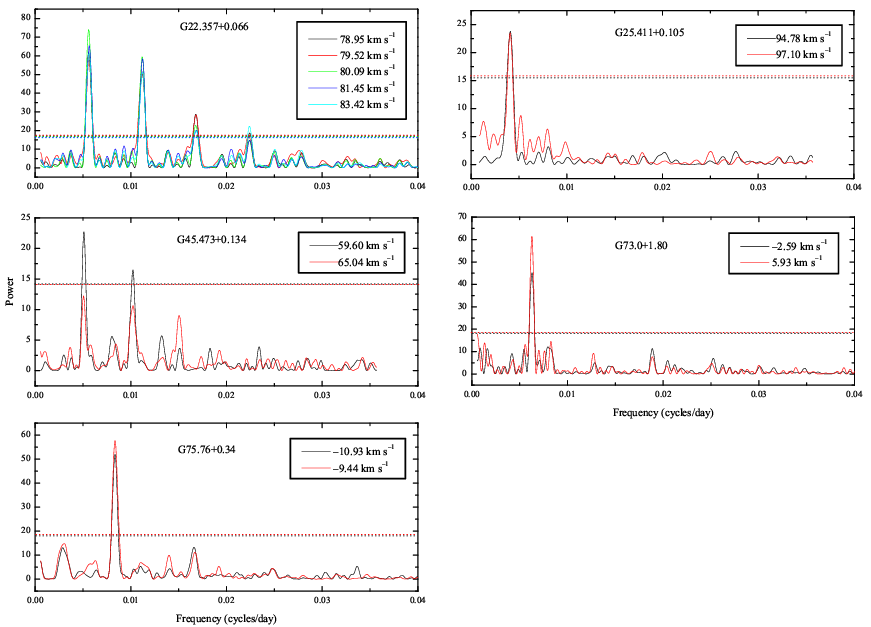}
\caption{The L-S periodograms for the features of the sources for significant frequencies, i.e. with a false 
alarm probability $\le10^{-4}$. The panels are labelled with the source name and the velocities of features
considered are given. The dotted horizontal lines mark a false-alarm probability of 0.01 per cent. 
Note harmonic series in the periodograms for the features of sources G22.357+0.066 and G45.473+0.134.
These significant frequencies seem unlikely to be real detections.
\label{periodograms}}
\end{figure*}
\end{appendix}


\end{document}